\documentclass[aps,pre,twocolumn,10pt,amsmath,amssymb,showpacs,floatfix,superscriptaddress]{revtex4-1}
\pdfoutput=1
\usepackage{overpic}
\usepackage[T1]{fontenc}
\usepackage{lmodern}
\usepackage{units}
\usepackage{color}
\usepackage{graphicx}
\usepackage{booktabs}
\usepackage{xcolor,graphicx}
\usepackage{multirow}

\newcommand{\ve}[1]{\ensuremath{\mbox{\boldmath$#1$}}}

\newcommand*{\ellk}{\ell_{\rm K}}
\newcommand*{\weff}{w_{\rm eff}}

\newcommand*{\fig}[1]{Fig.~\ref{fig:#1}}

\begin{document}
\title{Finite-size corrections for confined polymers in the extended de Gennes regime}
\author{T. St Clere Smithe} 
\author{V. Iarko} 
\affiliation{Department of Physics, University of Gothenburg, Sweden}
\author{A. Muralidhar}
\affiliation{Department of Chemical Engineering
and Materials Science, University of Minnesota -- Twin Cities,
421 Washington Avenue SE, Minneapolis, Minnesota 55455, USA}
\author{E. Werner}
\email{erik.werner@physics.gu.se}
\affiliation{Department of Physics, University of Gothenburg, Sweden}
\author{K. D. Dorfman}
\affiliation{Department of Chemical Engineering
and Materials Science, University of Minnesota -- Twin Cities,
421 Washington Avenue SE, Minneapolis, Minnesota 55455, USA}
\author{B. Mehlig}
\affiliation{Department of Physics, University of Gothenburg, Sweden}

\begin{abstract}
Theoretical results for the extension of a polymer confined to a
channel are usually derived in the limit of infinite contour
length. But experimental studies and simulations of DNA molecules
confined to nanochannels are not necessarily in this asymptotic limit.
We calculate the statistics of the span and the end-to-end distance 
of a semiflexible polymer of finite length in the extended de Gennes 
regime, exploiting the fact
that the problem can be mapped to a one-dimensional weakly
self-avoiding random walk. The results thus obtained compare favourably
with pruned-enriched Rosenbluth method (PERM) simulations of a
three-dimensional discrete wormlike chain model of DNA confined in a
nanochannel. We discuss the implications for experimental studies of
linear $\lambda$-DNA confined to nanochannels at the high ionic
strengths used in many experiments.
\end{abstract}
\maketitle

\section{Introduction}
The extension statistics of semiflexible polymers confined to channels
have been studied in great detail during the last years
\cite{Tree:2013a,Benkova2013,chen2013,chenYL2013,dai2013,dai2014b,
gupta2014,muralidhar2014,werner2014,Chen2014,werner2015,
alizadehheidari2015c,gupta2015,iarko2015}.
A large number of scaling laws have been identified, valid at
different values of the physical parameters describing the polymer and
the strength of confinement. (See Ref.~\cite{werner2015} for an
overview describing the different regimes of extensional fluctuations
of a polymer in a rectangular channel.) But these scaling laws are
strictly valid only in the limit where the contour length $L$ of the
polymer tends to infinity, and experiments and simulations
may not correspond to the asymptotic limit. Therefore it is important
to understand to what extent these results apply for finite contour
lengths, and how they must be corrected to accurately describe the
statistics obtained in experiments at finite values of $L$.

This motivated us to compute the finite-size corrections of the mean
values and variances of the end-to-end distance and the span of
confined polymers in the \lq extended de Gennes regime\rq{}
\cite{wang2011}. In
this regime we can make use of recent results \cite{werner2014} that
show how the conformational statistics of the confined polymer can be
mapped to a one-dimensional model. This allows us to analyse
finite-size corrections in two ways. First, a scaling law determines 
the magnitude of the finite-size corrections as a
function of the physical parameters. Second, Monte-Carlo simulations
of the one-dimensional model allow us to compute this scaling function
to a high degree of accuracy. We show that the finite-size effects
produced by this one-dimensional model provide a reasonably accurate
description of detailed three-dimensional simulations of a confined,
discrete wormlike chain model of DNA in a high ionic-strength buffer.
Finally we show that finite-size effects are of experimental
importance, by estimating the magnitude of the finite-size corrections
for the statistics of the end-to-end distance of
$\lambda$-DNA confined to nanochannels.

There are only limited data in the literature concerning finite-size
effects for confined DNA. In simulations, Muralidhar \emph{et al.}
\cite{muralidhar2014} estimated finite-size corrections for the
extension of DNA confined to square channels of widths between $30$ 
and $400$ nm at high ionic strength for contour lengths out to the 
asymptotic limit, corresponding to almost $10^4$ persistence lengths 
for the largest channel size. 
This analysis took advantage of pruned-enriched Rosenbluth method
simulations (PERM) \cite{grassberger1997,Prellberg2004} of a discrete
wormlike chain model consisting of a semiflexible chain of touching
beads. PERM is especially well suited to study finite-size effects, as
it natively produces data as a function of molecular weight while
avoiding the attrition problems due to excluded volume between
segments of the chain and between the chain and the walls.  
Muralidhar \emph{et al.} \cite{muralidhar2014} found the span of 
$\lambda$-DNA to be within 3\% of the value in the infinite-$L$ limit,
for the buffer strengths and channel dimensions they studied in this 
paper.

The study of finite-size effects is especially important in light of
recent work comparing experimental data on nanochannel-confined DNA to
simulations \cite{gupta2014,gupta2015} and theory
\cite{gupta2015,iarko2015}. Simulations and
theoretical predictions using reasonable values of the physical
properties of DNA are within around 10\% of the experimental
results. The remaining disagreement between theory/simulation and
experiment arises in part due to uncertainties in the 
theories used to estimate the persistence length 
\cite{Dobrynin:06,Hsieh:2008}
and effective width \cite{Stigter:77,iarko2015} of the polymer, which
are required inputs to the model, as well as experimental artifacts
that lead to molecular weight dispersity \cite{gupta2015}, such as
heterogeneous staining \cite{Nyberg:13}, photocleavage
\cite{Akerman:96}, or shear cleavage \cite{Kovacic:1995}, that may not
be controlled in all experiments. However, for $\lambda$-DNA the
finite-size effects are of a similar order of magnitude to the
aforementioned sources of error \cite{muralidhar2014}.
In the following we explain how the finite-size effects depend on the 
parameters describing the confined DNA.
As a result, we anticipate that the
results described below will prove very useful in future quantitative
comparisons between theory and experiment for DNA.

\section{Extended de Gennes regime}
For a semiflexible polymer of Kuhn length $\ellk$ and effective width
$\weff\ll \ellk$ confined to a square channel of cross section
$D\times D$, the extended de Gennes regime is defined by the
conditions \cite{odijk2008,wang2011,dai2013,dai2014b,werner2014}
\begin{equation}
\label{eqn:edG_Conditions}
\ellk \ll D \ll \ellk^2/\weff.
\end{equation}
In the extended de Gennes regime the mapping between the
polymer physics problem and the statistics of a weakly self-avoiding
random walk leads to exact predictions for the equilibrium
distribution of the span $R$ and the end-to-end distance $r$ of a
long polymer \cite{werner2014}. In its simplest form, the 
one-dimensional random-walk model is defined by two parameters: 
the total number $N$ of steps taken, and the energy penalty $\epsilon$
(measured in units of $k_{\rm B}T$). The mapping shows that in terms 
of the parameters of the confined polymer the energy penalty is given 
by \cite{werner2014}
\begin{equation}
\label{eq:epsilon}
\varepsilon = \frac{9\sqrt{3} \pi}{8} \beta,
\end{equation}
where the scaling factor $\beta$ is defined as 
\begin{align}
\beta &= \frac{\ellk \weff}{D^2}\,.
\label{eqn:scaling_polymer}
\end{align}

It follows from 
Eqs.~(\ref{eqn:edG_Conditions})-(\ref{eqn:scaling_polymer}) that 
$\varepsilon\ll 1$ in the extended de Gennes regime.
In this limit, results for this one-dimensional
model can be directly translated into predictions for the confined
semiflexible polymer if each step of the random walk is interpreted as
one Kuhn-length segment of the polymer, and the lattice spacing
of the one-dimensional model corresponds to a distance 
$\ellk/\sqrt{3}$ in the channel direction. Although in this paper we 
only consider macroscopic observables such as the end-to-end 
distance, 
we emphasise that the mapping holds also for microscopic observables, 
provided that one considers contour-length separations significantly 
larger than one Kuhn length.

In the limit $L\to \infty$, the first two moments of the equilibrium
distributions for $R$ and $r$ are identical and obey \cite{werner2014}
\begin{align}
\mu_{r} &= \mu_{R} 
= 0.9338(84)\, L\,\Big(\frac{\ellk \weff}{D^2}\Big)^{1/3} 
\label{eqn:extensionWLC}\,,\\
\sigma^2_{r} &= \sigma^2_{R} = 0.133(12)\, L \ellk\,. 
\label{eqn:varianceWLC}
\end{align}
The errors quoted in these equations reflect mathematical
bounds \cite{werner2014} derived from the exact results of
Ref.~\cite{vanderhofstad2003}.

The question posed in the Introduction is: how much do these
observables deviate from the above scaling predictions when $L$ is
finite? In the extended de Gennes regime the finite-size analysis is
simplified by a universal scaling relation \cite{werner2014} implying
that all observables must collapse onto a universal curve
when plotted as functions of the scaled variables
\begin{align}
L' = (L/\ellk) \beta^{2/3} , & \quad z' = (z / \ellk) \beta^{1/3}\,.
\label{eqn:mappingParameters_beta}
\end{align}
Here $L$ is the contour length, and $z$ stands for the distance 
between monomer locations in the channel direction ($z$-direction).
The scaling law derives from
a corresponding scaling law for the one-dimensional model in the limit 
$\varepsilon\to 0$: $n' = n
\varepsilon^{2/3}$ and $z' = z \varepsilon^{1/3}$ where $n$
is the number of steps and $z$ is the distance traversed 
by the random walk \cite{werner2014}.
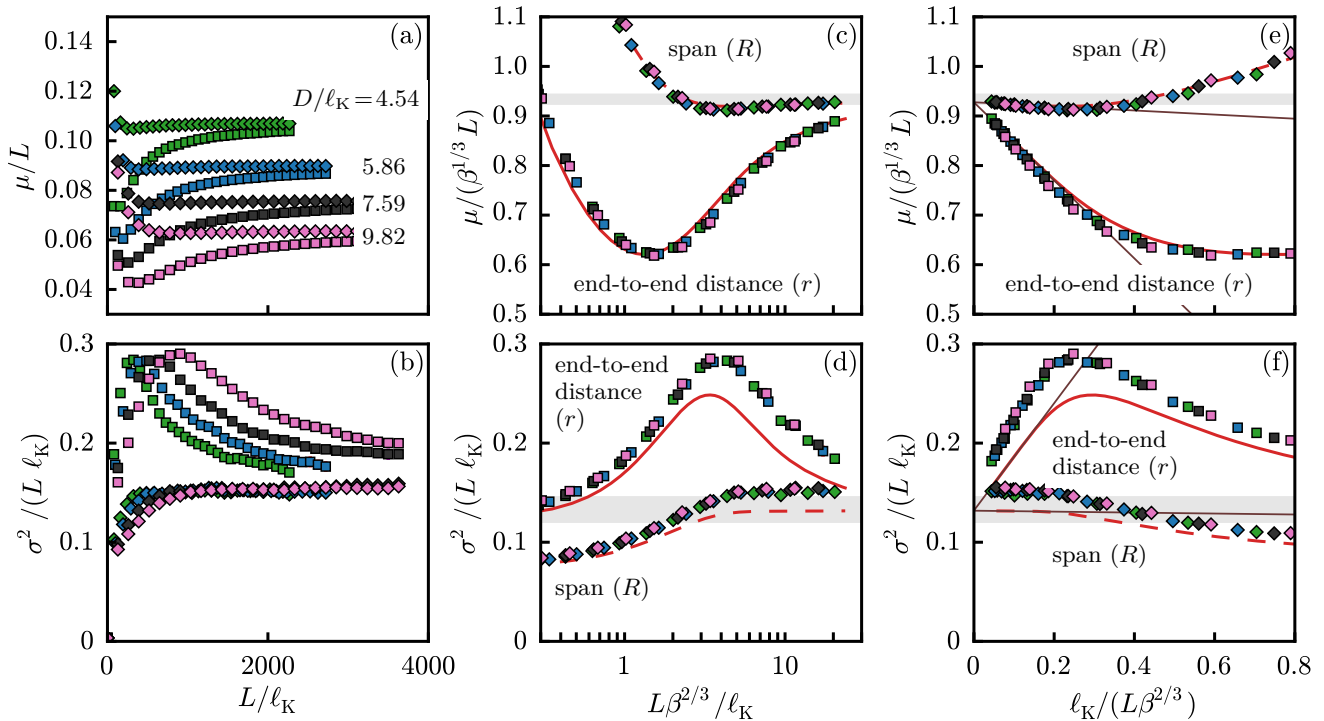
\begin{figure*}
\begin{overpic}[width=\textwidth]{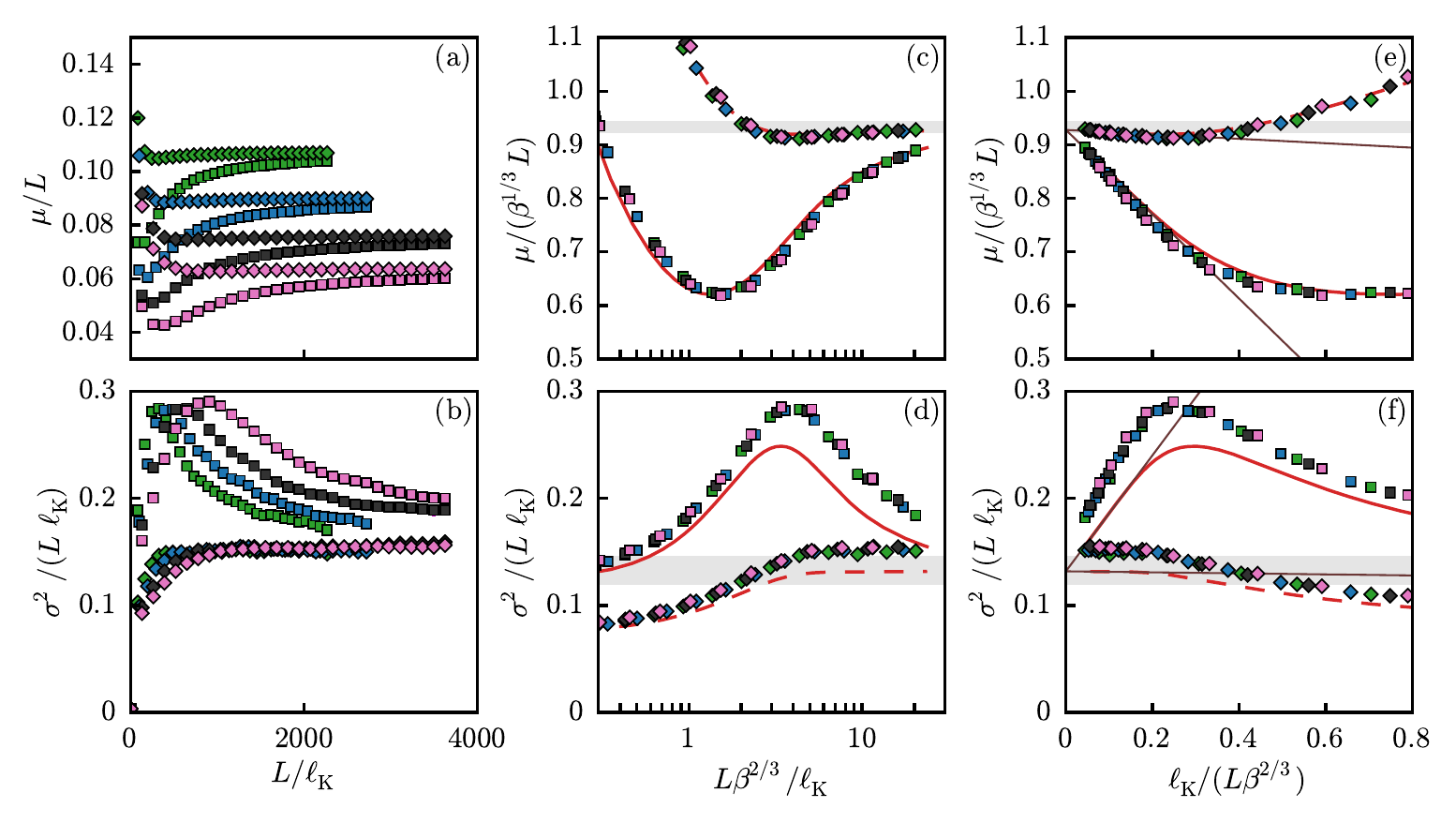}
\put(27.3,37.1){\colorbox{white}{$9.82$}}
\put(27.3,39.52){\colorbox{white}{$7.59$}}
\put(27.3,42.25){\colorbox{white}{$5.86$}}
\put(22.1,47.3){\colorbox{white}{$D/\ellk\!=\!4.54$}}
\put(50,51){\colorbox{white}{span ($R$)}}
\put(43,33.6){\colorbox{white}{end-to-end distance ($r$)}}
\put(41.6,11){\colorbox{white}{span ($R$)}}
\put(41.6,27.6){\colorbox{white}{end-to-end}}
\put(41.6,25.6){\colorbox{white}{distance }}
\put(41.6,23.6){\colorbox{white}{($r$)}}
\put(80,51){\colorbox{white}{span ($R$)}}
\put(75,33.6){\colorbox{white}{end-to-end distance ($r$)}}
\put(78.5,13.5){\colorbox{white}{span ($R$)}}
\put(78.5,22){\colorbox{white}{end-to-end}}
\put(78.5,20){\colorbox{white}{distance ($r$)}}
\end{overpic}
\caption{\label{fig:1d_results} {(a)}, {(b)} Mean and variance of the
  end-to-end distance $r$ and span $R$ of a discrete wormlike
  chain model of DNA in a high ionic-strength buffer ($d$ = 4.6 nm,
  $\ellk$ = 101.4 nm) confined in a channel of width $D$ as a function
  of contour length $L$. Symbols show results from three-dimensional
  simulations of the discrete wormlike chain model, for channel widths
  $D / \ellk = 9.82$ (pink), $D / \ellk = 7.59$ (grey), $D /
  \ellk = 5.86$ (blue) and $D / \ellk = 4.54$ (green, $\square$
  for $r$, $\diamond$ for $R$). {(c)},{(d)}: Same data, rescaled
  according to Eqs.~(\ref{eqn:scaling_polymer}) and
  (\ref{eqn:mappingParameters_beta}) with $\weff = d/1.45$ (see
  text). The grey shaded areas denote the asymptotic predictions of
  Eqs.~(\ref{eqn:extensionWLC}) and (\ref{eqn:varianceWLC}) in the
  limit $\beta\rightarrow 0$ and $L\rightarrow \infty$.
  Lines show the results from the one-dimensional model for a small
  value of $\varepsilon$ ($\varepsilon = 0.0011$). Solid lines for
  $r$, dashed for $R$. Panels (e) and (f) are as (c) and (d), except
  now as a function of $1/L$ to emphasise the linear dependence on
  $(L\beta^{2/3})^{-1}$ in the limit of $L \to \infty$ (thin lines
  show this extrapolation).
}
\end{figure*}

\section{Models and simulation method}

\subsection{One-dimensional self-avoiding random walk}

In order to determine how $\mu_{r}$, $\mu_{R}$, $\sigma^2_{r}$ and
$\sigma^2_{R}$ differ from the asymptotic results 
(\ref{eqn:extensionWLC}) and (\ref{eqn:varianceWLC}), we computed
these properties as functions of contour length using a chain-growth
method with a fixed number of chains at each contour length,
as described in Ref.~\cite{dai2014b}. The method can be summarised as 
follows.

The configuration for a given chain is generated by taking one step at
a time on a one-dimensional lattice, up to a maximum number of steps
$N$. A step is made with equal probability in either direction along
the lattice.  Random walks that revisit previously visited sites are
penalised as follows.  Assume that after step $j$ is made the random
walk lands on a site that has already been visited $\tau$ times (where
$\tau$ is known as the local time \cite{vanderhofstad2003}). The chain 
growth continues to
step $j+1$ with probability $p=\exp(-\varepsilon \tau)$ and the growth
is terminated at step $j$ with probability $1-p$.

Walks are generated in batches of constant size $N_c$, until a total
of 200 million walks are produced.
The initial statistical weight for each chain $k$ is then $w^{(k)}_1 =
1 / N_c$.  Each iteration consists of making one step for each walk in
the batch. The length-dependent averages and standard deviations for
the end-to-end distance are computed as the weighted values
\begin{eqnarray}
\mu_r(j) &=& Z_j^{-1} \sum_{k} w_j^{(k)} r_j^{(k)}, \label{eq:avg} \\
\sigma^2_r(j) &=& Z_j^{-1}\sum_k w_j^{(k)} [r_j^{(k)} - \mu_r(j)]^2, 
\label{eq:stdev}
\end{eqnarray}
with equivalent expressions for the span $R$ using all surviving
walks at step $j$. $Z_j = \sum_k w_j^{(k)}$ is the estimate of the
partition function after $j$ steps relative to the reference state.
If $N_p$ configurations are terminated at step $j$, then
we randomly select $N_p$ of the surviving chains and make copies of
these chains. All of the statistical weights are then updated as
\begin{equation}
w^{(k)}_j = w^{(k)}_{j-1} \left(\frac{N_c - N_p}{N_c} \right),
\end{equation}
Note that although the weights do not differ within
each batch under this re-weighting scheme, they do differ between
batches, and so the weights are used when computing the statistics
as in Eqs.~\eqref{eq:avg} and \eqref{eq:stdev}.

\subsection{Discrete wormlike chain model}

We also compare the results from the mapping of the weakly
self-avoiding random walk to detailed simulations of a discrete
wormlike chain confined in a nanochannel. With an eye towards
applications to experiments, we chose reasonable parameters for DNA at
high ionic strength, using $N + 1$ touching beads of diameter $d =
4.6$ nm, with contour length $L$ and Kuhn length $\ellk = 101.4$
nm. The centre of each bead is constrained to lie within a channel
with a square cross section of width $D$. Self-avoidance is imposed 
by disallowing
configurations where the distance between the centres of any two beads
is less than $d$.  Beads are connected by rigid bonds, and stiffness
is imposed by the bending potential
\begin{equation}
U(\theta_1,\ldots,\theta_{N}) 
= k_{\rm B} T \kappa \sum_{n=1}^{N-1} (1 - \cos \theta_n)\,,
\end{equation}
where $n$ is a bond index, $\theta_n$ is the angle between bonds $n$
and $n+1$, and $\kappa$ is the \lq{}bending constant\rq{}
\cite{muralidhar2014}. The Kuhn length is given by
\begin{equation}
\ellk = d \sum_{k=-\infty}^\infty \langle \ve t_n \ve t_{n+k}\rangle, 
\end{equation}
where $\ve t_n$ is the unit tangent vector to the chain backbone. 
This definition leads to the relationship
\cite{lodgeBook,Muralidhar:15}
\begin{equation}
\frac{\ellk}{d} 
= \frac{\kappa + \kappa \coth(\kappa) - 1}
{\kappa - \kappa \coth(\kappa) + 1}
\approx 2\kappa - 1   \,.
\end{equation}
The approximation is excellent for chains as stiff as the ones 
considered here.
Simulations of the discrete wormlike chain model use
the pruned-enriched Rosenbluth method (PERM) following the approach
described in detail in Refs.~\cite{tree2013,muralidhar2014}. The
simulation data here consist of 1.8 million tours in total for each
channel size.

\section{Results and Discussion }
Figure~\ref{fig:1d_results} presents numerical results of the
simulations described above. Panels (a) and (b) show the finite-size
corrections for the means and variances of the span and the
end-to-end distance for the discrete wormlike chain model of confined
DNA. We see that the magnitude of the finite-size corrections depends
not only on $L$ but also the channel size $D$ and on the observable in
question. In comparing the weakly self-avoiding random walk to the
detailed simulation data, it is important to keep in mind that
Eqs.~(\ref{eqn:extensionWLC}) to (\ref{eqn:mappingParameters_beta})
are valid for arbitrary polymer models provided that $\weff$ is {\it
defined} in terms of the excluded volume $v$ of a Kuhn length 
segment as $v = (\pi/2)\ellk^2 \weff$
\cite{werner2014}. This formula is valid for slender stiff
cylindrical rods of width $\weff\ll \ellk$
\cite{onsager1949}. But we cannot calculate $\weff$ because the 
excluded volume
$v$ is not known for the discrete wormlike chain model: the fact that 
the polymer is represented as a chain of touching beads complicates 
the geometrical analysis. A second complicating factor is that the 
Kuhn-length segments are not straight, this must affect the excluded 
volume. We therefore choose $\weff/d$ so that it gives the best fit 
between the one-dimensional model and the worm-like chain data for 
$\mu_r$ and $\mu_R$. We find $\weff/d =1/1.45$ for $D/\ellk = 5.61$ 
and use the same ratio for the other channel sizes.

Panels (c) and (d) of Fig.~\ref{fig:1d_results} demonstrate the
predicted universal form of the finite-size corrections.  We see that
by rescaling the data according to Eqs.~(\ref{eqn:scaling_polymer})
and (\ref{eqn:mappingParameters_beta}) the curves for different
physical parameters collapse onto one universal curve determined by
the one-dimensional random-walk model.

For the means the agreement is excellent, for the variances less so.
There are two plausible reasons for this. Firstly, since $D \gg \ellk$
holds only approximately for the smaller channel sizes used here 
($D/\ellk = 4.54,\; 5.86$), the polymer has a tendency to align with 
the channel which causes the variance to increase
\cite{werner2012}. Indeed, as we approach the small-channel side of
the extended de Gennes regime, this alignment effect and the
corresponding effect on the variance becomes very strong
\cite{muralidhar2014a}. 
Secondly, $D \ll \ellk^2 / \weff$ holds
only approximately for the bigger channel sizes 
($D/\ellk = 7.59,\; 9.82$ corresponds to 
$\ellk^2 /(D\weff) = 4.21,\; 3.25$), and the variance is expected to 
increase when this inequality is not strongly satisfied 
\cite{dai2013}. Thus, our results suggest that for DNA in a high 
ionic-strength buffer the modest separation of length scales between 
$\ellk$ and $\weff$ means that the measured variance agrees with the 
theoretical prediction (\ref{eqn:varianceWLC}) to within 10\% at best.

Panels (e) and (f) of Fig.~\ref{fig:1d_results} demonstrate that the
finite-size corrections are of order $L^{-1}$ in the limit of large
values of $L$ (straight lines in this limit). Extrapolation to
$L\to\infty$ yields results consistent with
Eqs.~(\ref{eqn:extensionWLC}) and (\ref{eqn:varianceWLC}), which are
indicated by the grey shaded regions of the figure.  Finally,
finite-size corrections to both the mean and the variance of the
span are seen to be much smaller than those for the end-to-end
distance, as noted in Ref.~\cite{muralidhar2014}.  In all cases we
observe that $\mu_r < \mu_R$, as expected. But for the variances the
relation is instead $\sigma_r > \sigma_R$. The latter observation
results from the form of the distributions for the end-to-end distance
and the span as the chain size increases. While both
distributions eventually converge for infinite chain lengths
\cite{muralidhar2014}, for finite chains the width of the distribution
for the end-to-end distance is wider than the distribution for the
span since the ends of the chains can fluctuate within the
terminal blob.

\begin{figure}
\includegraphics[width=\columnwidth]{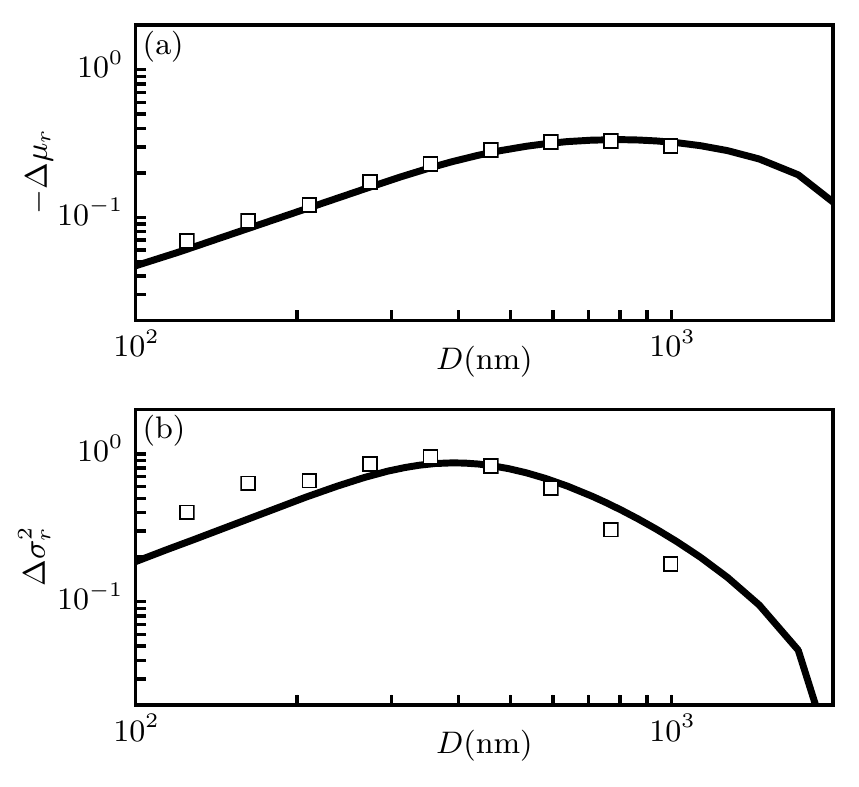}  
\caption{\label{fig:fss} Finite-size corrections in
  Eq.~(\ref{eqn:finiteSizeCorrection_general}) for the end-to-end
  distance $r$ for a model of $\lambda$-DNA in channels of different
  widths $D$.  Panel (a) shows the corrections for the mean $\mu_{r}$,
  and panel (b) shows the corrections for the variance $\sigma^2_{r}$.
  Results for discrete wormlike chain model describing $\lambda$-DNA
  ($d = 4.6$ nm, $\ellk = 101.4$ nm, $L$ = 16 $\mu$m) as a function of
  channel width $D$ (symbols).  Simulations of the one-dimensional
  random-walk model (solid lines). }
\end{figure}

We conclude our discussion by estimating how important finite-size
effects are for $\lambda$-DNA in the extended de Gennes regime. To
this end we define the finite-size correction $\Delta_X$ of an
extensive observable $X$ as
\begin{equation}
\label{eqn:finiteSizeCorrection_general}
  \Delta_X \equiv \frac{X/L}{\lim\limits_{L \to \infty} X/L} - 1\,.
\end{equation}
\fig{fss} shows the finite-size corrections for the mean (a) and the
variance (b) of the end-to-end distance $r$ for
$\lambda$-DNA ($L = 16\, \mu$m) as a
function of channel size $D$.  Symbols show the results of the
simulations of the three-dimensional discrete wormlike chain, using
the data in Fig.~\ref{fig:1d_results}(a,b), and
Eq.~(\ref{eqn:finiteSizeCorrection_general}). To estimate
$\lim\limits_{L\to\infty} X/L$ we used the mean and the variance of
the span for the largest values of $L$ available.  As mentioned
above, the mean and variance of the span converge more rapidly
but to the same limit as those of the end-to-end distance.  Results
for the one-dimensional random-walk model are shown as solid lines,
obtained from the red solid lines in Fig.~\ref{fig:1d_results}(c,d).
Using Eqs.~(\ref{eq:epsilon}) and (\ref{eqn:scaling_polymer}) we
convert $x$-axis values in Fig.~\ref{fig:1d_results}(c,d) to channel
width $D$, with $\varepsilon=0.0011$
and $\weff = d/1.45=3.17$ nm; note that one can use any value of
$\varepsilon \ll 1$ since there is a scaling relationship.  We 
estimate the limit $\lim\limits_{L\to\infty} X/L$ by extrapolation.

We note good agreement between the two models for the
mean of the end-to-end distance, with some disagreement for the
variance (discussed above). It is remarkable that there is good 
agreement between the theory and the simulations even for small values
of $D$, that correspond to the border of the extended de Gennes 
regime ($D \approx \ellk$).

We see that the finite-size corrections for the mean of the end-to-end
distance can be as large as $34$\%. The corresponding corrections for 
the variance are even larger, reaching 87\% for $D\approx 400$ nm.

We do not show the corresponding results for the span since the
finite-size corrections in this case are much smaller, and thus both
of less interest and more difficult to estimate reliably.

\section{Conclusions}
We have estimated finite-size corrections for the extension of a
confined semiflexible polymer of finite length $L$ in the extended de
Gennes regime. In this regime the problem can be mapped to a
one-dimensional weakly self-avoiding random walk, yielding
universal scaling relations that determine the finite-size
corrections. We confirm the scaling predictions by simulations of a
discrete wormlike chain model.

\begin{acknowledgments} 
Financial support from Vetenskapsr\aa{}det (grant number 2013-3992), 
from the G\"oran Gustafsson Foundation for Research in Natural 
Sciences and Medicine and the National Institutes of Health 
(R01-HG006851) is gratefully acknowledged. The computational work was 
carried out in part using computing resources at the University of 
Minnesota Supercomputing Institute.
\end{acknowledgments}


\begin{thebibliography}{31}%
\makeatletter
\providecommand \@ifxundefined [1]{%
 \@ifx{#1\undefined}
}%
\providecommand \@ifnum [1]{%
 \ifnum #1\expandafter \@firstoftwo
 \else \expandafter \@secondoftwo
 \fi
}%
\providecommand \@ifx [1]{%
 \ifx #1\expandafter \@firstoftwo
 \else \expandafter \@secondoftwo
 \fi
}%
\providecommand \natexlab [1]{#1}%
\providecommand \enquote  [1]{``#1''}%
\providecommand \bibnamefont  [1]{#1}%
\providecommand \bibfnamefont [1]{#1}%
\providecommand \citenamefont [1]{#1}%
\providecommand \href@noop [0]{\@secondoftwo}%
\providecommand \href [0]{\begingroup \@sanitize@url \@href}%
\providecommand \@href[1]{\@@startlink{#1}\@@href}%
\providecommand \@@href[1]{\endgroup#1\@@endlink}%
\providecommand \@sanitize@url [0]{\catcode `\\12\catcode `\$12\catcode
  `\&12\catcode `\#12\catcode `\^12\catcode `\_12\catcode `\%12\relax}%
\providecommand \@@startlink[1]{}%
\providecommand \@@endlink[0]{}%
\providecommand \url  [0]{\begingroup\@sanitize@url \@url }%
\providecommand \@url [1]{\endgroup\@href {#1}{\urlprefix }}%
\providecommand \urlprefix  [0]{URL }%
\providecommand \Eprint [0]{\href }%
\providecommand \doibase [0]{http://dx.doi.org/}%
\providecommand \selectlanguage [0]{\@gobble}%
\providecommand \bibinfo  [0]{\@secondoftwo}%
\providecommand \bibfield  [0]{\@secondoftwo}%
\providecommand \translation [1]{[#1]}%
\providecommand \BibitemOpen [0]{}%
\providecommand \bibitemStop [0]{}%
\providecommand \bibitemNoStop [0]{.\EOS\space}%
\providecommand \EOS [0]{\spacefactor3000\relax}%
\providecommand \BibitemShut  [1]{\csname bibitem#1\endcsname}%
\let\auto@bib@innerbib\@empty
\bibitem [{\citenamefont {Tree}\ \emph
  {et~al.}(2013{\natexlab{a}})\citenamefont {Tree}, \citenamefont {Wang},\ and\
  \citenamefont {Dorfman}}]{Tree:2013a}%
  \BibitemOpen
  \bibfield  {author} {\bibinfo {author} {\bibfnamefont {D.~R.}\ \bibnamefont
  {Tree}}, \bibinfo {author} {\bibfnamefont {Y.}~\bibnamefont {Wang}}, \ and\
  \bibinfo {author} {\bibfnamefont {K.~D.}\ \bibnamefont {Dorfman}},\
  }\href@noop {} {\bibfield  {journal} {\bibinfo  {journal} {Phys. Rev. Lett.}\
  }\textbf {\bibinfo {volume} {110}},\ \bibinfo {pages} {208103} (\bibinfo
  {year} {2013}{\natexlab{a}})}\BibitemShut {NoStop}%
\bibitem [{\citenamefont {Benkova}\ and\ \citenamefont
  {Cifra}(2013)}]{Benkova2013}%
  \BibitemOpen
  \bibfield  {author} {\bibinfo {author} {\bibfnamefont {Z.}~\bibnamefont
  {Benkova}}\ and\ \bibinfo {author} {\bibfnamefont {P.}~\bibnamefont
  {Cifra}},\ }\href@noop {} {\bibfield  {journal} {\bibinfo  {journal}
  {Biochem. Soc. Trans.}\ }\textbf {\bibinfo {volume} {41}},\ \bibinfo {pages}
  {625} (\bibinfo {year} {2013})}\BibitemShut {NoStop}%
\bibitem [{\citenamefont {Chen}(2013{\natexlab{a}})}]{chen2013}%
  \BibitemOpen
  \bibfield  {author} {\bibinfo {author} {\bibfnamefont {J.~Z.~Y.}\
  \bibnamefont {Chen}},\ }\href@noop {} {\bibfield  {journal} {\bibinfo
  {journal} {Macromolecules}\ }\textbf {\bibinfo {volume} {46}},\ \bibinfo
  {pages} {9837} (\bibinfo {year} {2013}{\natexlab{a}})}\BibitemShut {NoStop}%
\bibitem [{\citenamefont {Chen}(2013{\natexlab{b}})}]{chenYL2013}%
  \BibitemOpen
  \bibfield  {author} {\bibinfo {author} {\bibfnamefont {Y.-L.}\ \bibnamefont
  {Chen}},\ }\href@noop {} {\bibfield  {journal} {\bibinfo  {journal}
  {Biomicrofluidics}\ }\textbf {\bibinfo {volume} {7}},\ \bibinfo {pages}
  {054119} (\bibinfo {year} {2013}{\natexlab{b}})}\BibitemShut {NoStop}%
\bibitem [{\citenamefont {Dai}\ and\ \citenamefont {Doyle}(2013)}]{dai2013}%
  \BibitemOpen
  \bibfield  {author} {\bibinfo {author} {\bibfnamefont {L.}~\bibnamefont
  {Dai}}\ and\ \bibinfo {author} {\bibfnamefont {P.~S.}\ \bibnamefont
  {Doyle}},\ }\href@noop {} {\bibfield  {journal} {\bibinfo  {journal}
  {Macromolecules}\ }\textbf {\bibinfo {volume} {46}},\ \bibinfo {pages} {6336}
  (\bibinfo {year} {2013})}\BibitemShut {NoStop}%
\bibitem [{\citenamefont {Dai}\ \emph {et~al.}(2014)\citenamefont {Dai},
  \citenamefont {van~der Maarel},\ and\ \citenamefont {Doyle}}]{dai2014b}%
  \BibitemOpen
  \bibfield  {author} {\bibinfo {author} {\bibfnamefont {L.}~\bibnamefont
  {Dai}}, \bibinfo {author} {\bibfnamefont {J.}~\bibnamefont {van~der Maarel}},
  \ and\ \bibinfo {author} {\bibfnamefont {P.~S.}\ \bibnamefont {Doyle}},\
  }\href@noop {} {\bibfield  {journal} {\bibinfo  {journal} {Macromolecules}\
  }\textbf {\bibinfo {volume} {47}},\ \bibinfo {pages} {2445} (\bibinfo {year}
  {2014})}\BibitemShut {NoStop}%
\bibitem [{\citenamefont {Gupta}\ \emph {et~al.}(2014)\citenamefont {Gupta},
  \citenamefont {Sheats}, \citenamefont {Muralidhar}, \citenamefont {Miller},
  \citenamefont {Huang}, \citenamefont {Mahshid}, \citenamefont {Dorfman},\
  and\ \citenamefont {Reisner}}]{gupta2014}%
  \BibitemOpen
  \bibfield  {author} {\bibinfo {author} {\bibfnamefont {D.}~\bibnamefont
  {Gupta}}, \bibinfo {author} {\bibfnamefont {J.}~\bibnamefont {Sheats}},
  \bibinfo {author} {\bibfnamefont {A.}~\bibnamefont {Muralidhar}}, \bibinfo
  {author} {\bibfnamefont {J.~J.}\ \bibnamefont {Miller}}, \bibinfo {author}
  {\bibfnamefont {D.~E.}\ \bibnamefont {Huang}}, \bibinfo {author}
  {\bibfnamefont {S.}~\bibnamefont {Mahshid}}, \bibinfo {author} {\bibfnamefont
  {K.~D.}\ \bibnamefont {Dorfman}}, \ and\ \bibinfo {author} {\bibfnamefont
  {W.}~\bibnamefont {Reisner}},\ }\href {\doibase 10.1063/1.4879515} {\bibfield
   {journal} {\bibinfo  {journal} {J. Chem. Phys.}\ }\textbf {\bibinfo {volume}
  {140}},\ \bibinfo {pages} {214901} (\bibinfo {year} {2014})}\BibitemShut
  {NoStop}%
\bibitem [{\citenamefont {Muralidhar}\ \emph
  {et~al.}(2014{\natexlab{a}})\citenamefont {Muralidhar}, \citenamefont {Tree},
  \citenamefont {Wang},\ and\ \citenamefont {Dorfman}}]{muralidhar2014}%
  \BibitemOpen
  \bibfield  {author} {\bibinfo {author} {\bibfnamefont {A.}~\bibnamefont
  {Muralidhar}}, \bibinfo {author} {\bibfnamefont {D.~R.}\ \bibnamefont
  {Tree}}, \bibinfo {author} {\bibfnamefont {Y.}~\bibnamefont {Wang}}, \ and\
  \bibinfo {author} {\bibfnamefont {K.~D.}\ \bibnamefont {Dorfman}},\ }\href
  {\doibase 10.1063/1.4865965} {\bibfield  {journal} {\bibinfo  {journal} {J.
  Chem. Phys.}\ }\textbf {\bibinfo {volume} {140}},\ \bibinfo {pages} {084905}
  (\bibinfo {year} {2014}{\natexlab{a}})}\BibitemShut {NoStop}%
\bibitem [{\citenamefont {Werner}\ and\ \citenamefont
  {Mehlig}(2014)}]{werner2014}%
  \BibitemOpen
  \bibfield  {author} {\bibinfo {author} {\bibfnamefont {E.}~\bibnamefont
  {Werner}}\ and\ \bibinfo {author} {\bibfnamefont {B.}~\bibnamefont
  {Mehlig}},\ }\href {\doibase 10.1103/PhysRevE.90.062602} {\bibfield
  {journal} {\bibinfo  {journal} {Phys. Rev. E}\ }\textbf {\bibinfo {volume}
  {90}},\ \bibinfo {pages} {062602} (\bibinfo {year} {2014})}\BibitemShut
  {NoStop}%
\bibitem [{\citenamefont {Chen}\ \emph {et~al.}(2014)\citenamefont {Chen},
  \citenamefont {Lin}, \citenamefont {Chang},\ and\ \citenamefont
  {Lin}}]{Chen2014}%
  \BibitemOpen
  \bibfield  {author} {\bibinfo {author} {\bibfnamefont {Y.-L.}\ \bibnamefont
  {Chen}}, \bibinfo {author} {\bibfnamefont {Y.-H.}\ \bibnamefont {Lin}},
  \bibinfo {author} {\bibfnamefont {J.-F.}\ \bibnamefont {Chang}}, \ and\
  \bibinfo {author} {\bibfnamefont {P.-K.}\ \bibnamefont {Lin}},\ }\href@noop
  {} {\bibfield  {journal} {\bibinfo  {journal} {Macromolecules}\ }\textbf
  {\bibinfo {volume} {47}},\ \bibinfo {pages} {1199} (\bibinfo {year}
  {2014})}\BibitemShut {NoStop}%
\bibitem [{\citenamefont {Werner}\ and\ \citenamefont
  {Mehlig}(2015)}]{werner2015}%
  \BibitemOpen
  \bibfield  {author} {\bibinfo {author} {\bibfnamefont {E.}~\bibnamefont
  {Werner}}\ and\ \bibinfo {author} {\bibfnamefont {B.}~\bibnamefont
  {Mehlig}},\ }\href@noop {} {\bibfield  {journal} {\bibinfo  {journal} {Phys.
  Rev. E}\ }\textbf {\bibinfo {volume} {91}},\ \bibinfo {pages} {050601(R)}
  (\bibinfo {year} {2015})}\BibitemShut {NoStop}%
\bibitem [{\citenamefont {Alizadehheidari}\ \emph {et~al.}(2015)\citenamefont
  {Alizadehheidari}, \citenamefont {Werner}, \citenamefont {Noble},
  \citenamefont {Reiter-Schad}, \citenamefont {Nyberg}, \citenamefont
  {Fritzsche}, \citenamefont {Mehlig}, \citenamefont {Tegenfeldt},
  \citenamefont {Ambj{\"o}rnsson}, \citenamefont {Persson},\ and\ \citenamefont
  {Westerlund}}]{alizadehheidari2015c}%
  \BibitemOpen
  \bibfield  {author} {\bibinfo {author} {\bibfnamefont {M.}~\bibnamefont
  {Alizadehheidari}}, \bibinfo {author} {\bibfnamefont {E.}~\bibnamefont
  {Werner}}, \bibinfo {author} {\bibfnamefont {C.}~\bibnamefont {Noble}},
  \bibinfo {author} {\bibfnamefont {M.}~\bibnamefont {Reiter-Schad}}, \bibinfo
  {author} {\bibfnamefont {L.~K.}\ \bibnamefont {Nyberg}}, \bibinfo {author}
  {\bibfnamefont {J.}~\bibnamefont {Fritzsche}}, \bibinfo {author}
  {\bibfnamefont {B.}~\bibnamefont {Mehlig}}, \bibinfo {author} {\bibfnamefont
  {J.~O.}\ \bibnamefont {Tegenfeldt}}, \bibinfo {author} {\bibfnamefont
  {T.}~\bibnamefont {Ambj{\"o}rnsson}}, \bibinfo {author} {\bibfnamefont
  {F.}~\bibnamefont {Persson}}, \ and\ \bibinfo {author} {\bibfnamefont
  {F.}~\bibnamefont {Westerlund}},\ }\href {\doibase 10.1021/ma5022067}
  {\bibfield  {journal} {\bibinfo  {journal} {Macromolecules}\ }\textbf
  {\bibinfo {volume} {48}},\ \bibinfo {pages} {871} (\bibinfo {year}
  {2015})}\BibitemShut {NoStop}%
\bibitem [{\citenamefont {Gupta}\ \emph {et~al.}(2015)\citenamefont {Gupta},
  \citenamefont {Miller}, \citenamefont {Muralidhar}, \citenamefont {Mahshid},
  \citenamefont {Reisner},\ and\ \citenamefont {Dorfman}}]{gupta2015}%
  \BibitemOpen
  \bibfield  {author} {\bibinfo {author} {\bibfnamefont {D.}~\bibnamefont
  {Gupta}}, \bibinfo {author} {\bibfnamefont {J.~J.}\ \bibnamefont {Miller}},
  \bibinfo {author} {\bibfnamefont {A.}~\bibnamefont {Muralidhar}}, \bibinfo
  {author} {\bibfnamefont {S.}~\bibnamefont {Mahshid}}, \bibinfo {author}
  {\bibfnamefont {W.}~\bibnamefont {Reisner}}, \ and\ \bibinfo {author}
  {\bibfnamefont {K.~D.}\ \bibnamefont {Dorfman}},\ }\href@noop {} {\bibfield
  {journal} {\bibinfo  {journal} {ACS MacroLett.}\ }\textbf {\bibinfo {volume}
  {4}},\ \bibinfo {pages} {759} (\bibinfo {year} {2015})}\BibitemShut {NoStop}%
\bibitem [{\citenamefont {Iarko}\ \emph {et~al.}(2015)\citenamefont {Iarko},
  \citenamefont {Werner}, \citenamefont {Nyberg}, \citenamefont {M{\"u}ller},
  \citenamefont {Fritzsche}, \citenamefont {Ambj{\"o}rnsson}, \citenamefont
  {Beech}, \citenamefont {Tegenfeldt}, \citenamefont {Mehlig}, \citenamefont
  {Westerlund},\ and\ \citenamefont {Mehlig}}]{iarko2015}%
  \BibitemOpen
  \bibfield  {author} {\bibinfo {author} {\bibfnamefont {V.}~\bibnamefont
  {Iarko}}, \bibinfo {author} {\bibfnamefont {E.}~\bibnamefont {Werner}},
  \bibinfo {author} {\bibfnamefont {L.~K.}\ \bibnamefont {Nyberg}}, \bibinfo
  {author} {\bibfnamefont {V.}~\bibnamefont {M{\"u}ller}}, \bibinfo {author}
  {\bibfnamefont {J.}~\bibnamefont {Fritzsche}}, \bibinfo {author}
  {\bibfnamefont {T.}~\bibnamefont {Ambj{\"o}rnsson}}, \bibinfo {author}
  {\bibfnamefont {J.~P.}\ \bibnamefont {Beech}}, \bibinfo {author}
  {\bibfnamefont {J.~O.}\ \bibnamefont {Tegenfeldt}}, \bibinfo {author}
  {\bibfnamefont {K.}~\bibnamefont {Mehlig}}, \bibinfo {author} {\bibfnamefont
  {F.}~\bibnamefont {Westerlund}}, \ and\ \bibinfo {author} {\bibfnamefont
  {B.}~\bibnamefont {Mehlig}},\ }\href {http://arxiv.org/abs/1506.02241}
  {\bibfield  {journal} {\bibinfo  {journal} {{arXiv}:1506.02241}\ } (\bibinfo
  {year} {2015})}\BibitemShut {NoStop}%
\bibitem [{\citenamefont {Wang}\ \emph {et~al.}(2011)\citenamefont {Wang},
  \citenamefont {Tree},\ and\ \citenamefont {Dorfman}}]{wang2011}%
  \BibitemOpen
  \bibfield  {author} {\bibinfo {author} {\bibfnamefont {Y.}~\bibnamefont
  {Wang}}, \bibinfo {author} {\bibfnamefont {D.~R.}\ \bibnamefont {Tree}}, \
  and\ \bibinfo {author} {\bibfnamefont {K.~D.}\ \bibnamefont {Dorfman}},\
  }\href@noop {} {\bibfield  {journal} {\bibinfo  {journal} {Macromolecules}\
  }\textbf {\bibinfo {volume} {44}},\ \bibinfo {pages} {6594} (\bibinfo {year}
  {2011})}\BibitemShut {NoStop}%
\bibitem [{\citenamefont {Grassberger}(1997)}]{grassberger1997}%
  \BibitemOpen
  \bibfield  {author} {\bibinfo {author} {\bibfnamefont {P.}~\bibnamefont
  {Grassberger}},\ }\href@noop {} {\bibfield  {journal} {\bibinfo  {journal}
  {Phys. Rev. E}\ }\textbf {\bibinfo {volume} {56}},\ \bibinfo {pages} {3682}
  (\bibinfo {year} {1997})}\BibitemShut {NoStop}%
\bibitem [{\citenamefont {Prellberg}\ and\ \citenamefont
  {Krawczyk}(2004)}]{Prellberg2004}%
  \BibitemOpen
  \bibfield  {author} {\bibinfo {author} {\bibfnamefont {T.}~\bibnamefont
  {Prellberg}}\ and\ \bibinfo {author} {\bibfnamefont {J.}~\bibnamefont
  {Krawczyk}},\ }\href@noop {} {\bibfield  {journal} {\bibinfo  {journal}
  {Phys. Rev. Lett.}\ }\textbf {\bibinfo {volume} {92}},\ \bibinfo {pages}
  {120602} (\bibinfo {year} {2004})}\BibitemShut {NoStop}%
\bibitem [{\citenamefont {Dobrynin}(2006)}]{Dobrynin:06}%
  \BibitemOpen
  \bibfield  {author} {\bibinfo {author} {\bibfnamefont {A.~V.}\ \bibnamefont
  {Dobrynin}},\ }\href@noop {} {\bibfield  {journal} {\bibinfo  {journal}
  {Macromolecules}\ }\textbf {\bibinfo {volume} {39}},\ \bibinfo {pages} {9519}
  (\bibinfo {year} {2006})}\BibitemShut {NoStop}%
\bibitem [{\citenamefont {Hsieh}\ \emph {et~al.}(2008)\citenamefont {Hsieh},
  \citenamefont {Balducci},\ and\ \citenamefont {Doyle}}]{Hsieh:2008}%
  \BibitemOpen
  \bibfield  {author} {\bibinfo {author} {\bibfnamefont {C.-C.}\ \bibnamefont
  {Hsieh}}, \bibinfo {author} {\bibfnamefont {A.}~\bibnamefont {Balducci}}, \
  and\ \bibinfo {author} {\bibfnamefont {P.~S.}\ \bibnamefont {Doyle}},\
  }\href@noop {} {\bibfield  {journal} {\bibinfo  {journal} {Nano Lett.}\
  }\textbf {\bibinfo {volume} {8}},\ \bibinfo {pages} {1683} (\bibinfo {year}
  {2008})}\BibitemShut {NoStop}%
\bibitem [{\citenamefont {Stigter}(1977)}]{Stigter:77}%
  \BibitemOpen
  \bibfield  {author} {\bibinfo {author} {\bibfnamefont {D.}~\bibnamefont
  {Stigter}},\ }\href@noop {} {\bibfield  {journal} {\bibinfo  {journal}
  {Biopolymers}\ }\textbf {\bibinfo {volume} {16}},\ \bibinfo {pages} {1435}
  (\bibinfo {year} {1977})}\BibitemShut {NoStop}%
\bibitem [{\citenamefont {Nyberg}\ \emph {et~al.}(2013)\citenamefont {Nyberg},
  \citenamefont {Persson}, \citenamefont {{\AA}kerman},\ and\ \citenamefont
  {Westerlund}}]{Nyberg:13}%
  \BibitemOpen
  \bibfield  {author} {\bibinfo {author} {\bibfnamefont {L.}~\bibnamefont
  {Nyberg}}, \bibinfo {author} {\bibfnamefont {F.}~\bibnamefont {Persson}},
  \bibinfo {author} {\bibfnamefont {B.}~\bibnamefont {{\AA}kerman}}, \ and\
  \bibinfo {author} {\bibfnamefont {F.}~\bibnamefont {Westerlund}},\
  }\href@noop {} {\bibfield  {journal} {\bibinfo  {journal} {Nucleic Acids
  Res.}\ }\textbf {\bibinfo {volume} {41}},\ \bibinfo {pages} {e184} (\bibinfo
  {year} {2013})}\BibitemShut {NoStop}%
\bibitem [{\citenamefont {Akerman}\ and\ \citenamefont
  {Tuite}(1996)}]{Akerman:96}%
  \BibitemOpen
  \bibfield  {author} {\bibinfo {author} {\bibfnamefont {B.}~\bibnamefont
  {Akerman}}\ and\ \bibinfo {author} {\bibfnamefont {E.}~\bibnamefont
  {Tuite}},\ }\href@noop {} {\bibfield  {journal} {\bibinfo  {journal} {Nucleic
  Acids Res.}\ }\textbf {\bibinfo {volume} {24}},\ \bibinfo {pages} {1080}
  (\bibinfo {year} {1996})}\BibitemShut {NoStop}%
\bibitem [{\citenamefont {Kovacic}\ \emph {et~al.}(1995)\citenamefont
  {Kovacic}, \citenamefont {Comai},\ and\ \citenamefont
  {Bendich}}]{Kovacic:1995}%
  \BibitemOpen
  \bibfield  {author} {\bibinfo {author} {\bibfnamefont {R.~T.}\ \bibnamefont
  {Kovacic}}, \bibinfo {author} {\bibfnamefont {L.}~\bibnamefont {Comai}}, \
  and\ \bibinfo {author} {\bibfnamefont {A.~J.}\ \bibnamefont {Bendich}},\
  }\href@noop {} {\bibfield  {journal} {\bibinfo  {journal} {Nucleic Acids
  Res.}\ }\textbf {\bibinfo {volume} {23}},\ \bibinfo {pages} {3999} (\bibinfo
  {year} {1995})}\BibitemShut {NoStop}%
\bibitem [{\citenamefont {Odijk}(2008)}]{odijk2008}%
  \BibitemOpen
  \bibfield  {author} {\bibinfo {author} {\bibfnamefont {T.}~\bibnamefont
  {Odijk}},\ }\href@noop {} {\bibfield  {journal} {\bibinfo  {journal} {Phys.
  Rev. E}\ }\textbf {\bibinfo {volume} {77}},\ \bibinfo {pages} {060901}
  (\bibinfo {year} {2008})}\BibitemShut {NoStop}%
\bibitem [{\citenamefont {van~der Hofstad}\ \emph {et~al.}(2003)\citenamefont
  {van~der Hofstad}, \citenamefont {den Hollander},\ and\ \citenamefont
  {K\"{o}nig}}]{vanderhofstad2003}%
  \BibitemOpen
  \bibfield  {author} {\bibinfo {author} {\bibfnamefont {R.}~\bibnamefont
  {van~der Hofstad}}, \bibinfo {author} {\bibfnamefont {F.}~\bibnamefont {den
  Hollander}}, \ and\ \bibinfo {author} {\bibfnamefont {W.}~\bibnamefont
  {K\"{o}nig}},\ }\href@noop {} {\bibfield  {journal} {\bibinfo  {journal}
  {{Probability Theory and Related Fields}}\ }\textbf {\bibinfo {volume}
  {{125}}},\ \bibinfo {pages} {483} (\bibinfo {year} {2003})}\BibitemShut
  {NoStop}%
\bibitem [{\citenamefont {Hiemenz}\ and\ \citenamefont
  {Lodge}(2007)}]{lodgeBook}%
  \BibitemOpen
  \bibfield  {author} {\bibinfo {author} {\bibfnamefont {P.~C.}\ \bibnamefont
  {Hiemenz}}\ and\ \bibinfo {author} {\bibfnamefont {T.~P.}\ \bibnamefont
  {Lodge}},\ }\href@noop {} {\emph {\bibinfo {title} {Polymer Chemistry}}}\
  (\bibinfo  {publisher} {CRC Press},\ \bibinfo {address} {Boca Raton, FL},\
  \bibinfo {year} {2007})\BibitemShut {NoStop}%
\bibitem [{\citenamefont {Muralidhar}\ and\ \citenamefont
  {Dorfman}(2015)}]{Muralidhar:15}%
  \BibitemOpen
  \bibfield  {author} {\bibinfo {author} {\bibfnamefont {A.}~\bibnamefont
  {Muralidhar}}\ and\ \bibinfo {author} {\bibfnamefont {K.~D.}\ \bibnamefont
  {Dorfman}},\ }\href@noop {} {\bibfield  {journal} {\bibinfo  {journal}
  {Macromolecules}\ }\textbf {\bibinfo {volume} {48}},\ \bibinfo {pages} {2829}
  (\bibinfo {year} {2015})}\BibitemShut {NoStop}%
\bibitem [{\citenamefont {Tree}\ \emph
  {et~al.}(2013{\natexlab{b}})\citenamefont {Tree}, \citenamefont {Muralidhar},
  \citenamefont {Doyle},\ and\ \citenamefont {Dorfman}}]{tree2013}%
  \BibitemOpen
  \bibfield  {author} {\bibinfo {author} {\bibfnamefont {D.~R.}\ \bibnamefont
  {Tree}}, \bibinfo {author} {\bibfnamefont {A.}~\bibnamefont {Muralidhar}},
  \bibinfo {author} {\bibfnamefont {P.~S.}\ \bibnamefont {Doyle}}, \ and\
  \bibinfo {author} {\bibfnamefont {K.~D.}\ \bibnamefont {Dorfman}},\
  }\href@noop {} {\bibfield  {journal} {\bibinfo  {journal} {Macromolecules}\
  }\textbf {\bibinfo {volume} {46}},\ \bibinfo {pages} {8369} (\bibinfo {year}
  {2013}{\natexlab{b}})}\BibitemShut {NoStop}%
\bibitem [{\citenamefont {Onsager}(1949)}]{onsager1949}%
  \BibitemOpen
  \bibfield  {author} {\bibinfo {author} {\bibfnamefont {L.}~\bibnamefont
  {Onsager}},\ }\href@noop {} {\bibfield  {journal} {\bibinfo  {journal} {Ann.
  N. Y. Acad. Sci.}\ }\textbf {\bibinfo {volume} {51}},\ \bibinfo {pages} {627}
  (\bibinfo {year} {1949})}\BibitemShut {NoStop}%
\bibitem [{\citenamefont {Werner}\ \emph {et~al.}(2012)\citenamefont {Werner},
  \citenamefont {Persson}, \citenamefont {Westerlund}, \citenamefont
  {Tegenfeldt},\ and\ \citenamefont {Mehlig}}]{werner2012}%
  \BibitemOpen
  \bibfield  {author} {\bibinfo {author} {\bibfnamefont {E.}~\bibnamefont
  {Werner}}, \bibinfo {author} {\bibfnamefont {F.}~\bibnamefont {Persson}},
  \bibinfo {author} {\bibfnamefont {F.}~\bibnamefont {Westerlund}}, \bibinfo
  {author} {\bibfnamefont {J.~O.}\ \bibnamefont {Tegenfeldt}}, \ and\ \bibinfo
  {author} {\bibfnamefont {B.}~\bibnamefont {Mehlig}},\ }\href {\doibase
  10.1103/PhysRevE.86.041802} {\bibfield  {journal} {\bibinfo  {journal} {Phys.
  Rev. E}\ }\textbf {\bibinfo {volume} {86}},\ \bibinfo {pages} {041802}
  (\bibinfo {year} {2012})}\BibitemShut {NoStop}%
\bibitem [{\citenamefont {Muralidhar}\ \emph
  {et~al.}(2014{\natexlab{b}})\citenamefont {Muralidhar}, \citenamefont
  {Tree},\ and\ \citenamefont {Dorfman}}]{muralidhar2014a}%
  \BibitemOpen
  \bibfield  {author} {\bibinfo {author} {\bibfnamefont {A.}~\bibnamefont
  {Muralidhar}}, \bibinfo {author} {\bibfnamefont {D.~R.}\ \bibnamefont
  {Tree}}, \ and\ \bibinfo {author} {\bibfnamefont {K.~D.}\ \bibnamefont
  {Dorfman}},\ }\href {\doibase 10.1021/ma501687k} {\bibfield  {journal}
  {\bibinfo  {journal} {Macromolecules}\ }\textbf {\bibinfo {volume} {47}},\
  \bibinfo {pages} {8446} (\bibinfo {year} {2014}{\natexlab{b}})}\BibitemShut
  {NoStop}%
\end{thebibliography}
\end{document}